\documentclass[twocolumn,twoside,slac_two]{revtex4}
\usepackage{graphicx}
\usepackage{fancyhdr}
\begin{document}

\title{The effect of annealing on the RHESSI gamma-ray detectors}

%

\author{P. Veres}
\affiliation{E\"otv\"os University, Budapest, Bolyai Military University, Budapesti, Hungary}
\author{J. \v{R}\'{\i}pa}
\affiliation{Charles University, Faculty of Mathematics and Physics, Astronomical Institute,\\
Prague, Czech Republic}
\author{C. Wigger}
\affiliation{Kantonsschule Wohlen, Switzerland}
\begin{abstract}
The performance of nine RHESSI germanium detectors has been gradually deteriorating since its launch in 2002 because of radiation damage caused by passing through the Earth's radiation belts. 
To restore its former sensitivity, the spectrometer underwent an annealing procedure in November 2007. It, however, changed the 
RHESSI response and affected gamma-ray burst measurements, e.g., the hardness ratios and the spectral capabilities bellow approximately 100\,keV.

\end{abstract}

\maketitle
\thispagestyle{fancy}


\section{Introduction}
 The Ramaty High Energy Solar Spectroscopic Imager (RHESSI) \cite{ref01} (http://hesperia.gsfc.nasa.gov/hessi) 
 is primarily dedicated for studying solar physics in X-ray and gamma ray region.  It's spectrometer \cite{ref02} 
 consists of nine germanium detectors, which are, however, only lightly shielded and thus also allow omnidirectional 
 gamma-ray burst (GRB) detection (http://grb.web.psi.ch) \cite{ref03}. The energy range extends from       
 50 keV up to 17 MeV. The effective area reaches up to $200$ cm$^2$. With a field of view of about half of the sky, 
 RHESSI observes about 70 GRBs per year. It can detect all three populations of GRBs (long, short and intermediate)
 \cite{ref04,ref02a}.

\section{The Annealing procedure}
In November 2007 the spectrometer underwent a procedure called annealing which was hoped to restore it's sensitivity 
that had been gradually deteriorating because of radiation damage \cite{ref05}. 
It resided in heating up the germanium detectors to over $90^{\circ}{\rm C}$ for one week (operating temperature is about $90$ K)
\footnote{http://hesperia.gsfc.nasa.gov/hessi/news/jan\_16\_08.htm} \footnote{http://sprg.ssl.berkeley.edu/$\sim$tohban/nuggets/?page\-=article\&article\_id=69}. 
This procedure was successful only partly, because the low-energy response was not improved as well as the high-energy one. 
We have found that GRBs observed after the annealing have hardness ratio measurements systematically shifted to 
higher values that those observed before.

\section{The effect of annealing on high-energy indices}
We used two spectral models for fitting GRB spectra.
The Band function is of the form:
\begin{equation}\label{eq:band}
\frac{dN}{dE} \sim \left\{
\begin{array}{ll}
E^{-\alpha} \exp{\left(-\frac{E}{E_0}\right)} & \mbox{if } E \leq E_{\mbox{break}} \\
E^{-\beta} & \mbox{if } E \leq E_{\mbox{break}} 
\end{array}
\right.
\end{equation}

The other model is the cutoff power-law (CPL):
\begin{equation}\label{eq:cpl}
\frac{dN}{dE} \sim 
E^{-\alpha} \exp{\left(-\frac{E}{E_0}\right)} 
\end{equation}
which is basically the low-energy part of the Band function with $\beta=\infty$

\begin{table*}[t]
\caption{Spectral fits of cutoff power law and Band function of some selected GRBs.
The RHESSI off-axis angle for all these GRBs is near right angle.}
\centering
\begin{tabular}{ccccccc}

\hline
\\[-2.5ex]
GRB    &  model & data      &   $\alpha$    &    $\beta$    & $E_p$(keV) & $\chi^2_r$ \\
\hline\hline
       &        & Swift     & 1.37$\pm$0.02 &               &            &  1.27      \\
061121 & CPL    & Konus*    & 1.32$\pm$0.05 &               & 606$\pm$80 &  1.01      \\
       &        & RHESSI    & 1.37$\pm$0.10 &               & 532$\pm$57 &  1.01      \\
\hline
080607 & CPL    & Swift     & 1.15$\pm$0.03 &               &            &  0.70      \\
       &        & RHESSI    &\textbf{2.33$\pm$0.18}&       & 432$\pm$19 &  1.39      \\
\hline
080825 & Band   & Fermi     & 0.54$\pm$0.21 & 2.29$\pm$0.35 & 180$\pm$23 &  1.23       \\
       &        & RHESSI    &\textbf{5.36$\pm$0.86}& 2.92$\pm$0.57 & 256$\pm$25 &  0.80       \\
\hline
\\[-3.5ex]
       &        &           &               &               &            &\tiny*GCN 5837\\
\end{tabular}
\end{table*}

\section{The effect of annealing on the measured hardness ratios}
Here we present the evolution of the average hardness ratios and 
their relation to the annealing.
Figures 3 and 4 show the development of the average GRB RHESSI hardness ratios H21 and H32 over the years. H21 is a 
low energy ratio. It is the ratio of the GRB counts in the energy ranges (120 - 400)keV / (25 - 120) keV. 
H32 is a higher energy hardness ratio. It is the ratio of the counts in the ranges (400 - 1500) keV / (120 - 400) keV. 
Also the development of the average GRB T90 durations is shown (the plotted errors are 2 sigma). 
Emphasised are the data after the annealing realised in Nov 2007.

\begin{figure}
\includegraphics[width=\columnwidth]{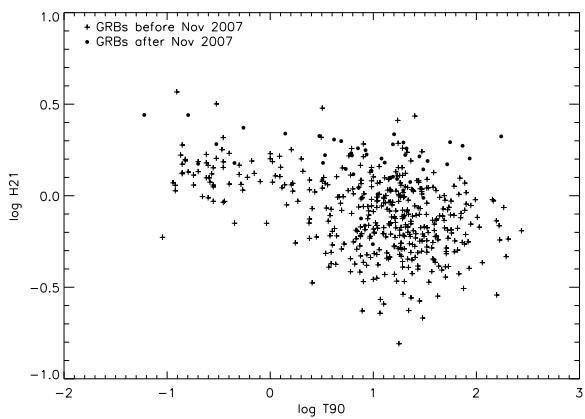}
\caption{The lower energy hardness ratio potted as function of the duration. Points after the annealing 
are systematically harder.}
\label{l2ea4-f1}
\end{figure}
\begin{figure}
\includegraphics[width=\columnwidth]{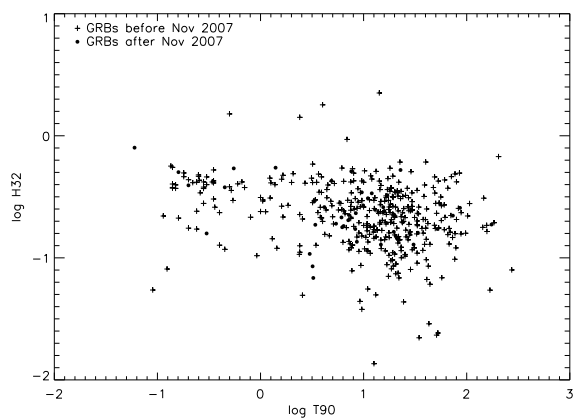}
\caption{The higher energy hardness ratio potted again in function of the duration. The effect of annealing is less evident.}
\label{l2e4-f1}
\end{figure}
\begin{figure}
\includegraphics[width=\columnwidth]{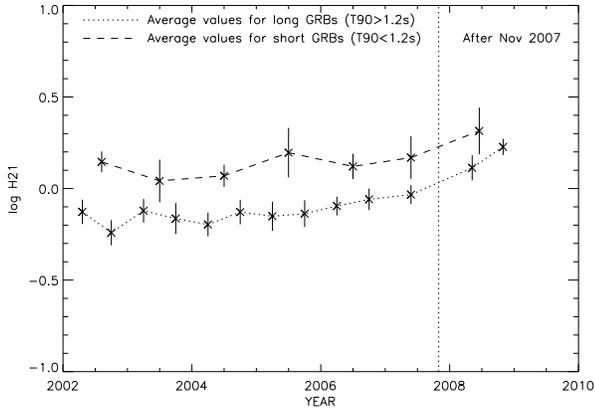}
\caption{The evolution of the lower energy hardness ratio in time for short- and long bursts. The vertical line 
marks the time of the annealing. An increasing trend is clearly seen.}
\label{l2e-f1}
\end{figure}
\begin{figure}
\includegraphics[width=\columnwidth]{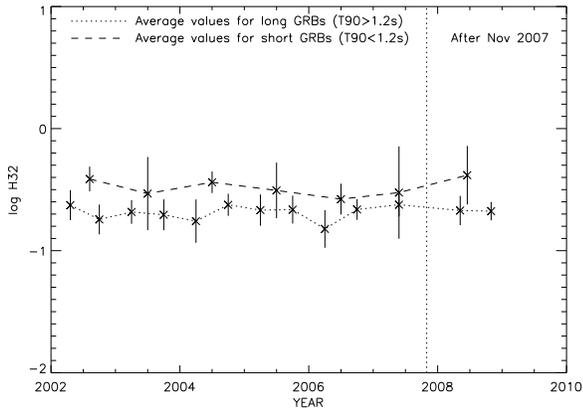}
\caption{The evolution of the higher energy hardness ratio in time for short- and long bursts. There is no significant
difference between the pre- and the post annealing phase.}
\label{l2f1}
\end{figure}
\begin{figure}
\includegraphics[width=\columnwidth]{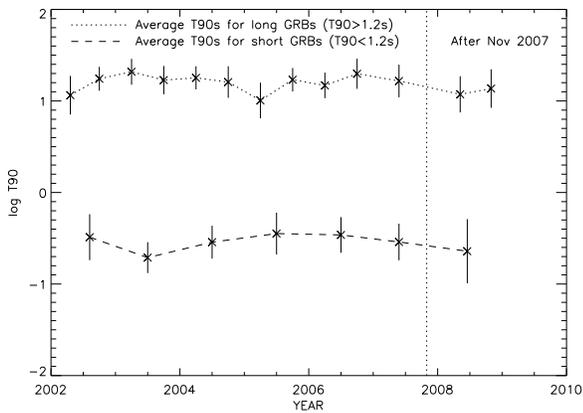}
\caption{Average $T_{90}$ for long- and short GRBs.}
\label{l}
\end{figure}

\section{Results and conclusion}
From the figures it is seen that the observed GRB low energy hardness ratios H21 were systematically shifted to higher 
values after the RHESSI annealing in Nov 2007. Contrary to this the high energy ratio H32 remains, on average, the same. 
The $T_{90}$ durations had not been affected.
   
   We have also compared the spectral parameters of 3 GRBs. Whereas the low-energy photon index for the pre-annealing burst 
   061121,  detected by Swift, Konus, and RHESSI, was found to be approximately the same, for the post-annealing bursts 
   the situation is different. The RHESSI low-energy index for bursts 080607 and 080825 markedly differ from 
   those obtained by the Swift or Fermi satellites.
   
   This finding and the H21 systematic shift point to the fact that the RHESSI low energy sensitivity was not recovered 
   well by the annealing procedure and using the RHESSI data for a future GRB spectral analysis, employing current 
   response matrix, might be problematic.

\begin{acknowledgments}
        This research is supported by Hungarian OTKA grant K077795 (P. V.), 
		by the GAUK grant No. 46307, by the GACR 205/08/H005 grant and by the 
        Research Program MSM0021620860 of the Ministry of Education of the Czech Republic (J. \v{R}.).
\end{acknowledgments}



\begin{thebibliography}{}
\bibitem{ref01} Lin, R.P., et al. 2002, Solar Physics, 210, 3
\bibitem{ref02} Smith, D.M., et al. 2002, Solar Physics, 210, 33
\bibitem{ref03} Smith, D.M., et al. 2003, Proceedings of SPIE Vol. 4851, 1163
\bibitem{ref04} {\v R}{\'{\i}}pa, J., et al. 2009, Astronomy \& Astrophysics, 498, 399
\bibitem{ref02a} Horv\'ath, I. et al. 2006, Astronomy \& Astrophysics, 447, 23

\bibitem{ref05} Bellm, E.~C., et al. 2008, AIP Conf. Proc., 1000, 154


\end{thebibliography}
\end{document}